\documentclass[%
 aip,
 amsmath,amssymb,
 reprint,%
]{revtex4-1}

\usepackage{graphicx}
\usepackage{dcolumn}
\usepackage{bm}
\usepackage[mathlines]{lineno}

\usepackage[utf8]{inputenc}
\usepackage[T1]{fontenc}
\usepackage{mathptmx}
\usepackage{etoolbox}

\usepackage{color}
\definecolor{Blue}{rgb}{0,0,1}
\definecolor{Red}{rgb}{0,0,0}
\definecolor{Green}{rgb}{0,1,0}

\usepackage{siunitx} 
\DeclareSIUnit\decibel{dB}
\DeclareSIUnit\fluxquanta{$\Phi_0$}
\DeclareSIUnit\ohm{$\Omega$}
\makeatletter
\def\@email#1#2{%
 \endgroup
 \patchcmd{\titleblock@produce}
  {\frontmatter@RRAPformat}
  {\frontmatter@RRAPformat{\produce@RRAP{*#1\href{mailto:#2}{#2}}}\frontmatter@RRAPformat}
  {}{}
}%
\makeatother
\begin{document}

\preprint{AIP/123-QED}

\title{Millikelvin Nb nanoSQUID-embedded tuneable resonator fabricated with a neon focused-ion-beam}
\author{Jamie A. Potter}
\affiliation{National Physical Laboratory, Hampton Road, Teddington, TW11 0LW, UK}

\author{Laith Meti}
\affiliation{National Physical Laboratory, Hampton Road, Teddington, TW11 0LW, UK}
\affiliation{London Centre for Nanotechnology, 17-19 Gordon Street, London, WC1H 0AH, UK}

\author{Gemma Chapman}
\affiliation{National Physical Laboratory, Hampton Road, Teddington, TW11 0LW, UK}

\author{Ed Romans}
\affiliation{London Centre for Nanotechnology, 17-19 Gordon Street, London, WC1H 0AH, UK}

\author{John Gallop}
\affiliation{National Physical Laboratory, Hampton Road, Teddington, TW11 0LW, UK}

\author{Ling Hao}
\affiliation{National Physical Laboratory, Hampton Road, Teddington, TW11 0LW, UK}
\email{ling.hao@npl.co.uk}

\date{\today}

\begin{abstract}
    SQUID-embedded superconducting resonators are of great interest due to their potential for coupling highly scalable superconducting circuits with quantum memories based on solid-state spin ensembles. Such an application requires a high-$Q$, frequency-tuneable resonator which is both resilient to magnetic field, and able to operate at millikelvin temperatures. These requirements motivate the use of a higher $H_{c}$ metal such as niobium, however the challenge then becomes to sufficiently reduce the operating temperature. We address this by presenting a monolithic Nb nanoSQUID-embedded resonator, where neon focused-ion-beam fabrication of the nanoSQUID results in a device displaying frequency tuneability at $T = 16$ mK. In order to assess the applicability of the device for coupling to small spin clusters, we characterise the flux sensitivity as a function of microwave drive power and externally applied magnetic field, and find that the noise is dominated by dielectric noise in the resonator. Finally, we discuss improvements to the device design which can dramatically improve the flux sensitivity, which highlights the promise of Nb SQUID-embedded resonators for hybrid superconductor-spin applications.
\end{abstract}

\maketitle

\let\clearpage\relax

A critical component of a full scale quantum computer is a quantum memory, allowing storage and on-demand retrieval of quantum information. \cite{Devoret_2013, Naik_2017} One of the most promising schemes to realise such a storage register is an ensemble of paramagnetic spins in a solid-state material, coupled to a resonant superconducting circuit. \cite{Wesenberg_2009,Morton_2018} Such a hybrid superconductor-spin system takes advantage of the long coherence times available in solid-state spin systems, as well as the flexibility and fast operation time of a superconducting quantum processor.

The longest coherence times in spin ensemble quantum memories are achieved by taking advantage of so-called clock transitions, where the electron spin strongly hybridises with a nuclear spin and results in an electron spin transition frequency that is first-order insensitive to fluctuations of magnetic field. \cite{Bollinger_1985} A number of material platforms have been studied in the context of clock transitions, for example bismuth donors in silicon (Si:Bi), \cite{Mohammady_2010,George_2010,Wolfowicz_2013,Mortemousque_2014,Ranjan_2020} rare-earth-ion-doped crystals, \cite{Zhong_2015,Ortu_2018} and molecular spins. \cite{Shiddiq_2016} In Si:Bi, four different clock transitions can be found at moderate magnetic field (<\SI{200}{\milli\tesla}), with transition frequencies in the \SIrange{5}{7.5}{\giga\hertz} range, and coherence times in excess of seconds have been demonstrated. \cite{Tyryshkin_2012,Wolfowicz_2013} 

The nature of the spin system sets a number of requirements for the superconducting resonator. In order to work at a Si:Bi clock transition, the resonator must be resilient to magnetic field, and operable at a sufficiently low temperature so as to minimise thermal occupation of higher electron spin states, \cite{Albanese_2020} which is typically satisfied for $T<\SI{50}{\milli\kelvin}$. \cite{Wesenberg_2009} In addition, the strong coupling regime is achieved when the coupling strength $g$ between the ensemble and the resonator is greater than both the spin linewidth \textit{and} the resonator linewidth, necessitating a high quality-factor resonator. \cite{Morton_2018}

\begin{figure}[b]
    \centering
    \includegraphics[width=\columnwidth]{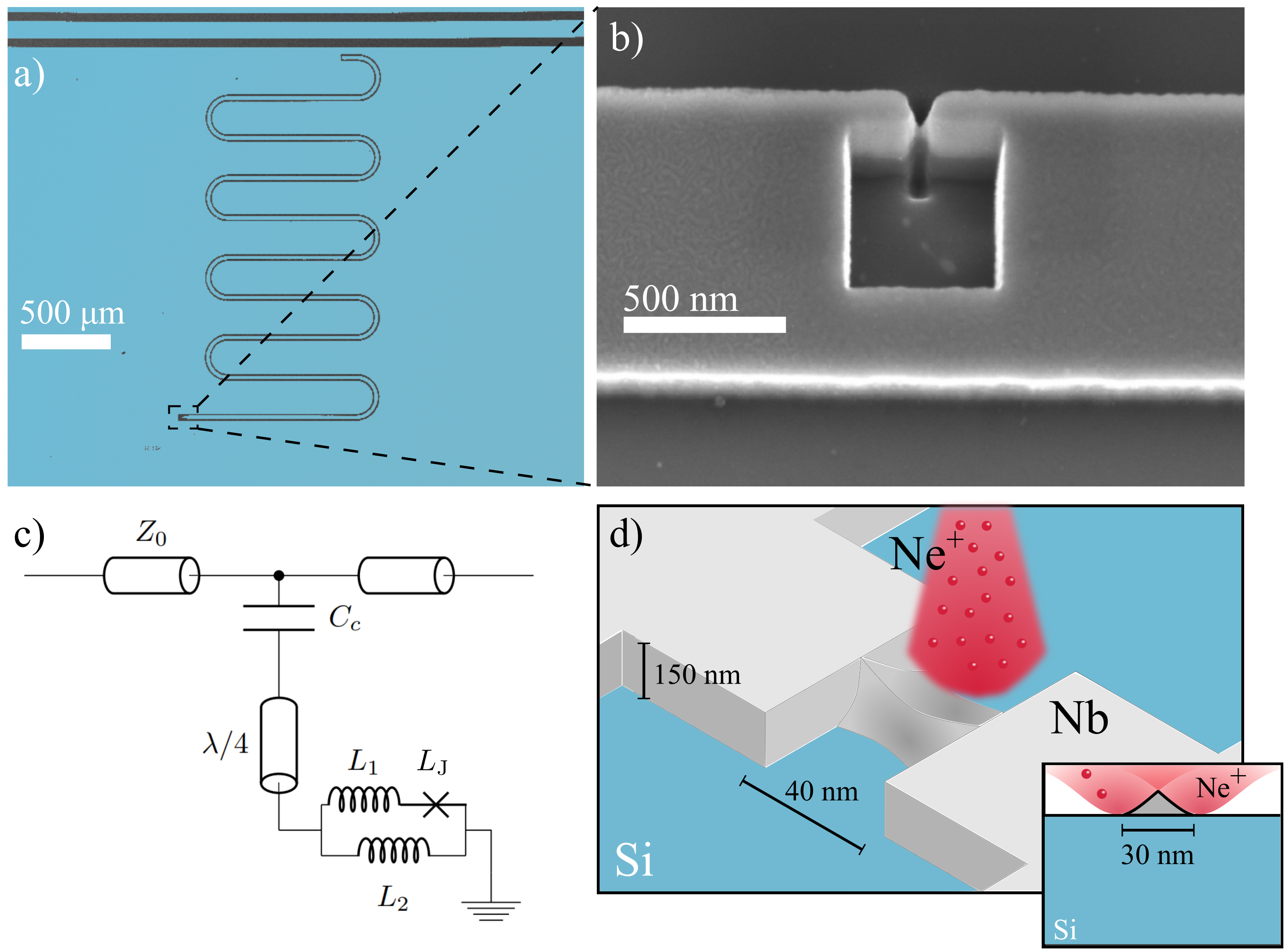}
    \caption{(a) False-coloured SEM image of $\lambda/4$ CPW resonator, capacitively coupled to a common microwave feedline (blue\textemdash{}niobium film). The resonator is terminated to ground via an rf-SQUID, shown in (b). (c) Circuit diagram of the flux-tuneable resonator. (d) Illustration of the formation of a 3d nanobridge with the neon FIB. The beam has a finite spread, with intensity decreasing outwards from the centre. This leads to a bridge which is thinner than the surrounding film, and is wider at the base and narrower towards the top.}
    \label{fig:device}
\end{figure}

\begin{figure*}[t]
    \centering
    \includegraphics[width=0.9\textwidth]{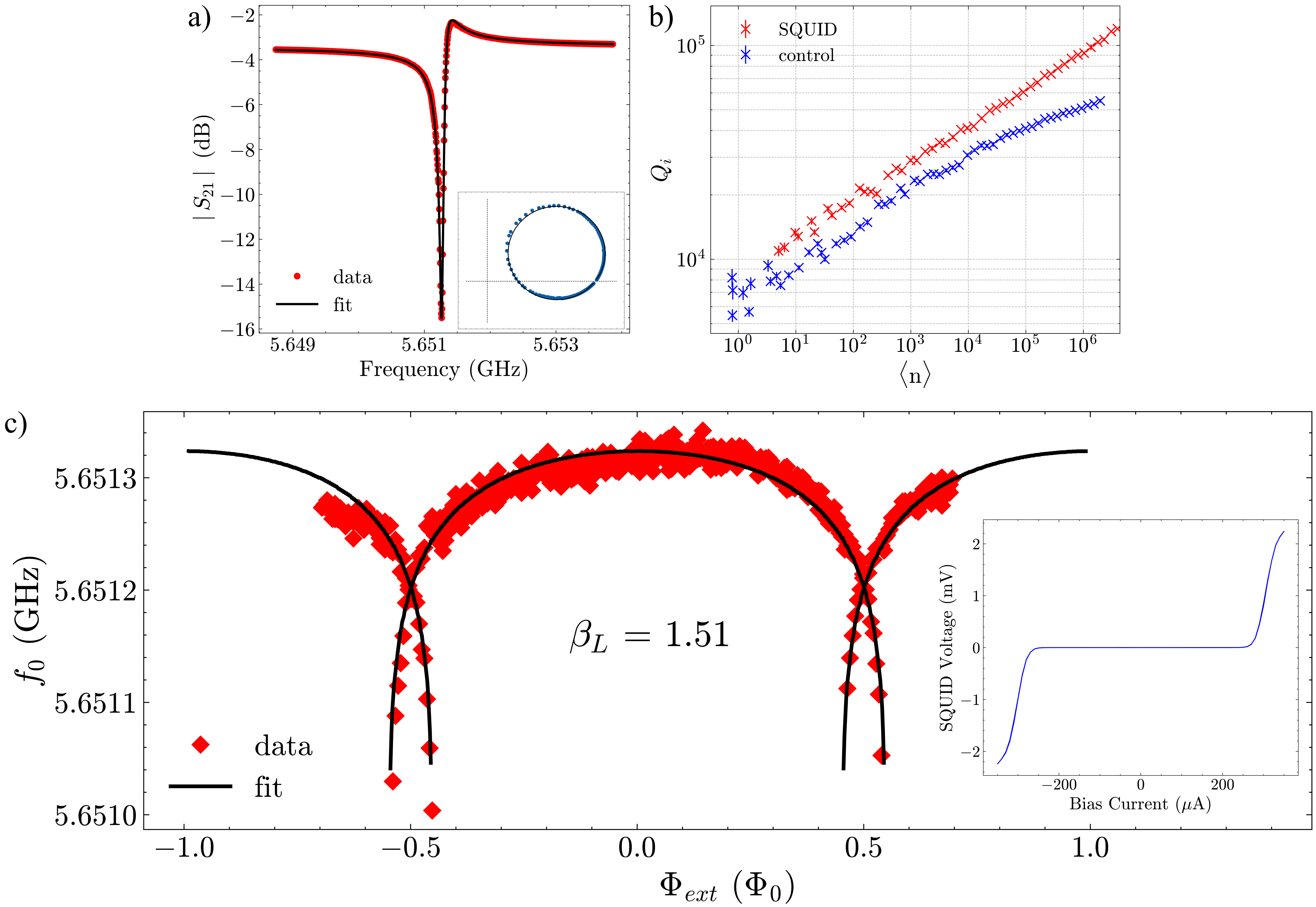}
    \caption{(a) $S_{21}$ amplitude response (red points) of the SQUID-embedded resonator at $P_{in} = -80$ dBm and $T = 16$ mK. Inset: $S_{21}$ data in the complex plane, along with analytic circle fit (black line). (b) Internal $Q$-factor as a function of average photon number for the SQUID-embedded resonator compared with a control resonator on the same chip. Vertical error bars show the statistical error which is estimated by the fitting algorithm. \cite{Probst_2015} (c) Tuning of the resonance in response to an externally applied magnetic flux, the value of which is inferred from the periodicity of the tuning. Points are measured data, the solid line is a fit from which we extract an inductive screening parameter of ${\beta_{L}=2\pi L_{\mathrm{loop}}I_{0}/\Phi_{0}=1.51}$. Inset - Current-voltage characteristic of a reference dc SQUID with similar nanobridges, measured at $T = \SI{16}{\milli\kelvin}$.}
    \label{fig:resonator}
\end{figure*}

It is also desirable to be able to bring the resonator in and out of resonance with the spin ensemble, in order to take advantage of Purcell-enhanced spin relaxation. \cite{Bienfait_2016_1} One way to achieve this tuneability is by including a SQUID or array of SQUIDs into the resonator, which act as a flux-tuneable inductance, enabling fast modulation of the resonant frequency. A number of SQUID-tuneable resonators have been demonstrated operating at $T<\SI{50}{\milli\kelvin}$; \cite{Palacios-Laloy_2008,Sandberg_2008,Levenson-Falk_2011,Stockklauser_2017} however, all of these cases use Al/AlO$_{x}$/Al Josephson junctions or aluminium nanobridge weak links, both of which are expected to suffer from poor magnetic field resilience. It has also been found that aluminium is not a suitable material for coupling to spins in silicon, due to strain-induced broadening of the spin transition resulting from a mismatch in the coefficient of thermal expansion between aluminium and silicon. \cite{Bienfait_2016_2,Pla_2018,O'Sullivan_2020} An alternative approach would be to use a material like niobium. \cite{Xu_2023,Uhl_2024} Niobium nanoSQUIDs have been studied extensively in the dc regime, \cite{Hasselbach_2002,Lam_2003,Troeman_2007,Gallop_2015,Granata_2016,Rodrigo_2020, Polychroniou_2020} demonstrating excellent noise performance, \cite{Hao_2008} and coupling to nanoscale magnetic systems. \cite{Hao_2011,Gallop_2016} However, the lowest operation temperature yet achieved in this material\cite{Kennedy_2019} is \SI{300}{\milli\kelvin}, which is not sufficient to fully suppress thermal excitation of a coupled spin system. The challenge in reducing the working temperature further is the necessary reduction of the nanobridge dimensions, required because the Ginzburg-Landau coherence length $\xi_{\mathrm{GL}}(T)$ gets shorter with decreasing temperature. In order to exhibit Josephson-like behaviour, a nanobridge must satisfy the criterion \cite{Likharev_1979} that ${l\lesssim3.5\xi_{\mathrm{GL}}(T)}$. In fact, the true length of a nanobridge extends some way into the electrodes, giving rise to an effective length which is greater than the geometric length. We estimate the coherence length of niobium at \SI{20}{\milli\kelvin} to be approximately \SI{30}{\nano\metre}, assuming that $\xi_{\mathrm{GL}}(T)\propto(1-\frac{T}{T_{\mathrm{c}}})^{-1/2}$ and $\xi_{\mathrm{GL}}\approx\SI{40}{\nano\metre}$ at \SI{4.2}{\kelvin}. Therefore we require a nanobridge with effective length less than \SI{105}{\nano\metre}.

In this article we present an rf nanoSQUID-embedded resonator fabricated entirely from a single layer of Nb, exhibiting high $Q$-factor and flux tuneability at $T=\SI{16}{\milli\kelvin}$, thus satisfying the set of requirements for strong coupling to an ensemble of Bi donor spins in Si. The low operation temperature and high $Q$-factor is enabled by a high-resolution neon focused-ion-beam (FIB) fabrication method. We go on to characterise the flux sensitivity of the device, in order to assess its suitability for coupling to single spins in close proximity to the nanobridge, for example for high-sensitivity electron-spin resonance (ESR). \cite{Bienfait_2016_2,Probst_2017}

The device is shown in Fig.~\ref{fig:device}. A $\lambda/4$ coplanar waveguide (CPW) resonator is patterned from a 150 nm-thick niobium film, deposited by dc magnetron sputtering on a high-resistivity silicon substrate. The CPW central conductor is \SI{10}{\micro\metre} wide, and the separation to the ground plane is chosen to give ${Z_{0}\approx50\text{ }\Omega}$. The resonator is coupled to a common ${50\text{ }\Omega}$ feedline via a capacitance $C_{c}\approx\SI{0.7}{\femto\farad}$. At the short-circuit terminated end of the resonator, the centre conductor tapers to a width of \SI{1}{\micro\metre} to allow for the inclusion of a nanoSQUID.

\begin{figure*}[t]
    \centering
    \includegraphics[width=0.9\textwidth]{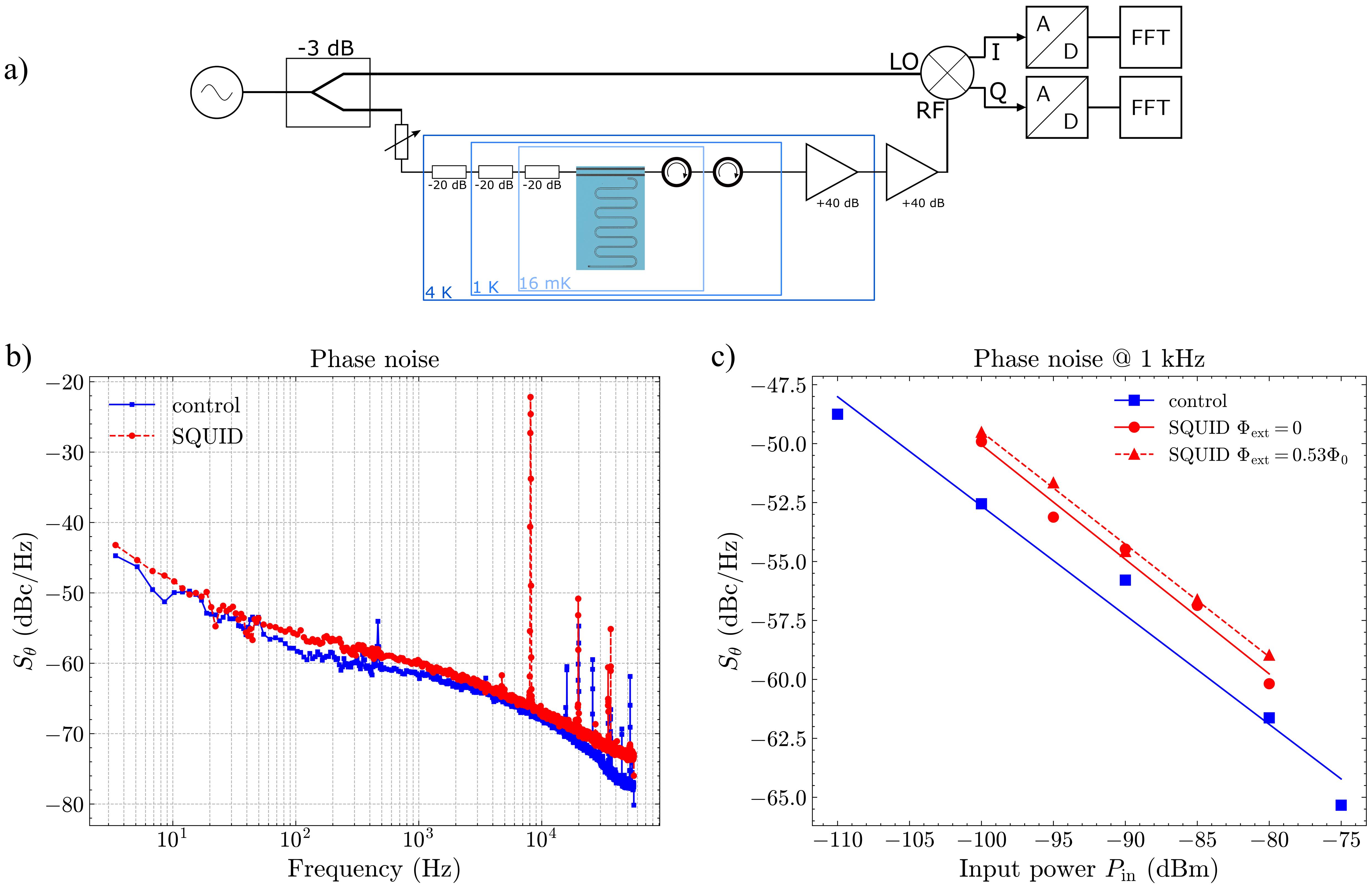}
    \caption{(a) Homodyne measurement setup for characterising the resonator noise. (b) Noise spectra in the phase quadrature, for the control resonator (blue) and the SQUID-embedded resonator (red). Both noise spectra are obtained at zero flux with an input power of -80 dBm. (c) Phase noise at 1 kHz as a function of the input power. Straight line fits show that $S_{\theta}\propto P_{\mathrm{in}}^{-0.5}$ for both devices.}
    \label{fig:noisepower}
\end{figure*}

After the microwave circuitry has been fabricated by patterning with \SI{100}{\kilo\electronvolt} electron-beam lithography (EBL) and reactive-ion etching with a CHF$_{3}$ and SF$_{6}$ plasma, the nanoSQUID is added by Ne FIB milling. The Ne FIB boasts a beam profile as small as \SI{2}{\nano\metre} and excellent depth of focus, making it the ideal tool for high resolution patterning of relatively thick films. Using an acceleration voltage of \SI{20}{\kilo\electronvolt} and a dose of \SI{2.5}{\nano\coulomb\per\micro\metre\tothe{2}}, we cut a ${\SI{500}{\nano\metre}\times\SI{500}{\nano\metre}}$ square hole into the tapered track, and then add a single narrow constriction. Using a scanning-electron microscope (SEM), we measure the constriction to be \SI{30}{\nano\metre} wide by \SI{40}{\nano\metre} long. SRIM modelling of \SI{20}{\kilo\electronvolt} Ne ions in Nb metal suggests a lateral implantation depth of approximately 25 nm. Therefore, we expect that most or all of the nanobridge is probably implanted with Ne, reducing the superconducting gap below that of pure Nb. We also stress here that the transverse dimensions of the nanobridge are not uniform along its length [see Fig.1(b)] and the additional damage induced by FIB milling will clearly influence both the electron mean free path and the crystallinity of the material within the bridge. Thus, it is difficult to give an accurate estimate of the effective coherence length of the bridge material. We also note that the yield of these \SI{}{\milli\kelvin} nanobridges is modest \textemdash{} around 50\% \textemdash{} which is a result of the nanobridge dimensions being comparable to the limit of  the Ne FIB patterning resolution.

SEM imaging of the nanoSQUID [Fig.~\ref{fig:device}(b)] reveals another consequence of the Ne FIB milling\textemdash{} that the constriction is significantly thinner than the surrounding film. Based on a tilted SEM image, we estimate the nanobridge to be approximately \SI{40}{\nano\metre} thick at its centre. This type of structure is referred to as a 3d nanobridge, and it has been shown \cite{Vijay_2009,Levenson-Falk_2016} to better constrain the nonlinear effect to within the geometric length of the nanobridge, thus decreasing the effective length. It has also been shown to provide greater critical current modulation \cite{Bouchiat_2001,Vijay_2010, Chen_2016} and nonlinearity. \cite{Uhl_2024} In addition to this, the reduced cross-sectional area of the bridge will naturally result in a lower critical current, which is desirable for operating at lower temperatures without thermal hysteresis. \cite{Skocpol_1974} Earlier reports of 3d nanobridges \cite{Vijay_2010, Chen_2016} have used multi-layer deposition approaches, which significantly increases the fabrication complexity, as well as introducing challenges such as ensuring good contact between layers and avoiding unwanted oxides. Our method is similar to that recently reported in Ref.\citenum{Uhl_2024}, with the distinction that the 3d profile of the bridge arises from the finite spread of the beam [as illustrated in Fig.~\ref{fig:device}(d)], requiring fewer patterning steps.

Following fabrication, the sample was mounted to the base plate of a dilution refrigerator with a base temperature of \SI{16}{\milli\kelvin}. The sample was positioned inside the bore of a superconducting solenoid capable of supplying a maximum magnetic field of \SI{2}{\tesla}. The input microwave line was attenuated by a total of ${-60\text{ dB}}$ to thermalise the incoming signals to the sample temperature, and the output line contains a cryogenic low-noise high-electron-mobility transistor (HEMT) amplifier at \SI{3}{\kelvin} and two further low-noise amplifiers at room temperature. Noise propagating backwards from the first amplifier input was attenuated by two cryogenic isolators providing a total reverse isolation of ${-40\text{ dB}}$.

As a first step of device characterisation, we measure the $S_{21}$ transmission through the feedline with a vector network analyser (VNA). Fig.~\ref{fig:resonator}(a) shows the zero-field resonance at an applied microwave power $P_{in}$ of approximately -80 dBm. From the fit, \cite{Probst_2015} we find a resonant frequency of ${f_{0} = \SI{5.6513}{\giga\hertz}}$ and an internal $Q$-factor of $Q_{i} = 1.41\times10^{5}$. Fig.~\ref{fig:resonator}(b) shows how $Q_{i}$ scales with the average photon number in the resonator $\langle n \rangle$, compared with a bare control resonator on the same chip. $\langle n \rangle$ is inferred using the following equation: \cite{Geaney_2019}
\begin{equation}
    \langle n \rangle = \frac{\langle E_{\mathrm{int}} \rangle}{hf_{0}} = \frac{2}{\pi}\frac{(Q_{i}^{-1}+Q_{c}^{-1})^{2}}{Q_{c}^{2}}\frac{P_{in}}{hf_{0}^{2}},
\end{equation}
where $Q_{c}$ is the coupling $Q$-factor. Both resonators have approximately the same $Q$, reaching a single-photon $Q_{i}$ of $7\times10^{3}$. This shows that the presence of the SQUID introduces no appreciable losses into the circuit, an observation consistent with previous studies of Ne-FIB-fabricated constrictions in CPW resonators. \cite{Burnett_2017, Kennedy_2019}

Fig.~\ref{fig:resonator}(c) shows the response of the device to a globally applied magnetic flux. We observe hysteretic tuning of the resonant frequency with a maximum range of \SI{300}{\kilo\hertz}. As the flux is increased from zero, the resonant frequency shifts downwards until an applied flux of \SI{0.545}{\fluxquanta}, at which point $f_{0}$ undergoes a discontinuous jump onto a new curve. From the flux at which the jump occurs we are able to extract ${\beta_{L}=2\pi L_{\mathrm{loop}}I_{0}/\Phi_{0}=1.51}$. We used the software 3D-MLSI \cite{Khapaev_2001} to calculate the inductance of the SQUID loop to be ${L_{\mathrm{loop}}=\SI{1.55}{\pico\henry}}$, which gives a nanobridge critical current of ${I_{0}=\SI{320}{\micro\ampere}}$. For comparison, the inset of Fig.~\ref{fig:resonator}(c) shows the current-voltage characteristic of a dc SQUID at $T = \SI{16}{\milli\kelvin}$, showing non-hysteretic behaviour, and a critical current of \SI{470}{\micro\ampere}, corresponding to a single nanobridge critical current of \SI{235}{\micro\ampere}.

In order to quantitatively model the flux dependence, we consider the device as a lossless transmission line with characteristic impedance $Z_{0}$, terminated by a load impedance $Z_{L}$. \cite{Potter_2023} The input impedance of the circuit is
\begin{equation} \label{eq:Zin}
    Z_{in}(f,Z_{L}) = Z_{0}\frac{Z_{L}+iZ_{0}\tan{\big(\frac{2\pi fl}{c}}\big)}{Z_{0}+iZ_{L}\tan{\big(\frac{2\pi fl}{c}}\big)},
\end{equation}
where $c=1/\sqrt{L_{l}C_{l}}$ is the speed of light in the resonator. $L_{l}$ and $C_{l}$ are respectively the inductance and capacitance per unit length of the bare CPW, and are calculated analytically using a conformal mapping technique. \cite{Simons_2004} We assume the load impedance arising from the SQUID is purely inductive, and thus $Z_{L}=i2\pi fL(\Phi_{\mathrm{tot}})$, where $\Phi_{\mathrm{tot}}$ is the total flux in the SQUID loop, and is related to the externally applied flux $\Phi_{\mathrm{ext}}$ via the following equation:
\begin{equation} \label{eq:Phitot}
    \frac{\Phi_{\mathrm{tot}}}{\Phi_{0}} = \frac{\Phi_{\mathrm{ext}}}{\Phi_{0}} - \frac{\beta_{L}}{2\pi}\sin{\bigg(2\pi\frac{\Phi_{\mathrm{tot}}}{\Phi_{0}}\bigg)}.
\end{equation}
The microwave currents in the resonator flow across the SQUID loop, rather than circulating around it, so the relevant inductance $L$ is a parallel combination of the inductance of the junction branch $L_{1}+L_{J}$ and non-junction branch $L_{2}$ of the SQUID loop
\begin{equation} \label{eq:L}
    \frac{1}{L} = \frac{1}{L_{1}+\frac{L_{J}(0)}{|\cos{(\pi\Phi_{\mathrm{tot}}/\Phi_{0})}|}} + \frac{1}{L_{2}},
\end{equation}
where $L_{1}+L_{2}=L_{\mathrm{loop}}$ is the total geometric inductance of the loop, and $L_{J}$ is the Josephson inductance of the nanobridge. For a notch type resonator, the imaginary part of the input impedance goes to infinity at resonance, therefore $f_{0}$ is calculated by numerically determining the drive frequency at which this condition is met. The black lines in Fig.~\ref{fig:resonator}(c) show the fit of this model to the experimental data.

We then characterised the noise of both the SQUID-embedded and the control resonator using a homodyne detection scheme. The methodology follows that presented in Ref.\citenum{Gao_2007}, and the measurement setup is shown in Fig.~\ref{fig:noisepower}(a). A microwave synthesiser is used to probe the resonator with a carrier frequency within a narrow range around $f_{0}$. The signal transmitted through the feedline is amplified and then demodulated by mixing with a copy of the original signal using an $IQ$ mixer. The output voltages $I$ and $Q$ are sampled by a two-channel digital oscilloscope over a 10 s interval with a sample rate of 112 kS/s, and then the pair of traces $[I,Q]$ is scaled and rotated \cite{Mazin_2004} in order to eliminate changes in amplitude and phase due to the measurement circuit. As the drive frequency $f$ is varied, the mean value of the transformed voltages $\langle[I,Q]^{T}\rangle$ traces out a circle in the $IQ$ plane. At the resonant frequency $f_{0}$, one finds that the fluctuations about the mean occur primarily in the direction tangent to the resonance circle. This corresponds to fluctuations in the phase of the resonator's electric field, which can be roughly understood as jitter in the resonant frequency. Finally, the power spectral density of these phase fluctuations can be calculated to allow spectral analysis of the resonator phase noise.

\begin{figure}[t]
    \centering
    \includegraphics[width=\columnwidth]{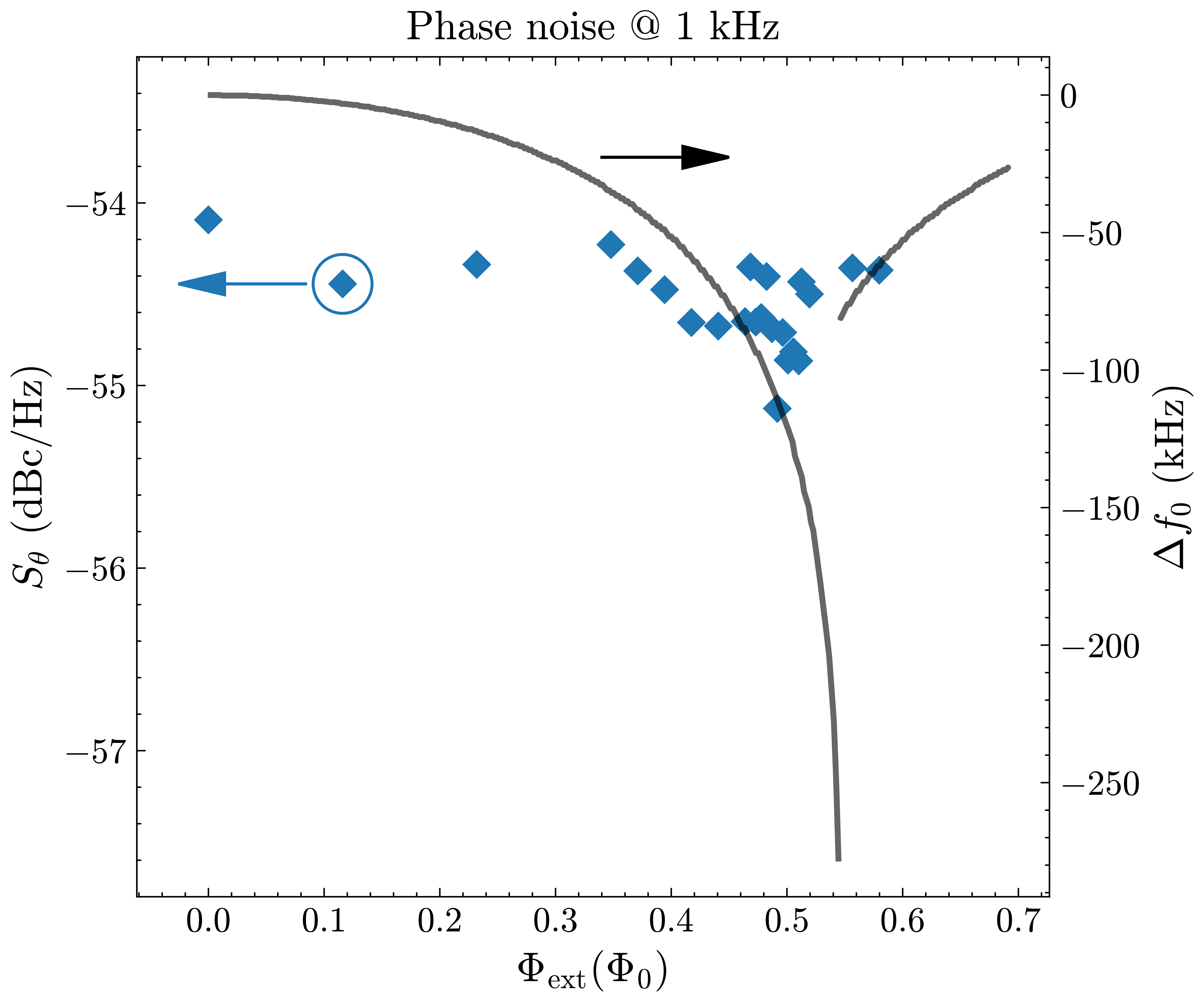}
    \caption{Phase noise of the SQUID-embedded resonator as a function of applied magnetic flux $\Phi_{\mathrm{ext}}$, with $P_{in}=-85$ dBm. The datapoints (left axis) show phase noise $S_{\theta}$. The black line (right axis) replicates the black line from Fig.~\ref{fig:resonator}(c), which is the fit to the frequency tuning of the resonator as the external flux is ramped upwards from zero. This is included to highlight the flux at which the resonant frequency exhibits a discontinuity. Note that the same discontinuity is not evident in the phase noise data.}
    \label{fig:noiseflux}
\end{figure}

Fig.~\ref{fig:noisepower}(b) shows phase noise spectra $S_{\theta}$ of the control and the SQUID-embedded resonator, at a drive power of -80 dBm and zero applied magnetic field. For $f<\SI{100}{\hertz}$, we observe a $f^{-1}$ dependence, and we see a roll-off at frequencies greater than \SI{10}{\kilo\hertz} which is related to the bandwidth of the resonator. In the region in between, there is an $f^{-0.5}$ dependence. There have been several reports of $S_{\theta}\propto f^{-0.5}$ in similar devices; \cite{Day_2003,Baselmans_2007,Gao_2007,Kumar_2008} however, it has been highlighted \cite{Burnett_2013,Faoro_2015} that this functional dependence cannot be explained by any theory of the underlying noise mechanisms, and that the true behaviour should follow $S_{\theta}\propto f^{-1}$. Instead, the $f^{-0.5}$ dependence of the spectrum is likely due to insufficient sampling time, which in our case was limited by the available memory depth of our digitiser. Nevertheless, the method we have used here will allow for consistent comparison between our devices, and examination of the dependence of resonator noise on drive power and magnetic field.

Both the control and the SQUID-embedded resonator noise exhibit a power dependence of $P^{-0.5}$ [Fig.~\ref{fig:noisepower}(c)], which is predicted by the generalised tunneling model of two-level systems (TLS) interacting with the resonator. \cite{Faoro_2015} It is generally accepted that the dominant source of noise in planar superconducting resonators arises from changes in the effective dielectric constant due to a fluctuating bath of TLS, and the fact that the SQUID-embedded resonator follows exactly the same power dependence suggests that the origin of the noise is the same for this device. We also observe in Fig.~\ref{fig:noisepower}(c) that there is no difference in the noise of the SQUID resonator whether the SQUID is flux-biased at its most flux-sensitive point or not. We proceeded to examine the flux dependence of the SQUID-embedded resonator by measuring the resonator noise as a function of applied magnetic field, with $P_{in}=-85$ dBm, and the result is shown in Fig. 4. We find that the resonator phase noise is independent of the applied flux. The conclusion from this is that the intrinsic SQUID noise is significantly lower than the measured resonator noise. We note here that SQUID flux noise and resonator phase noise can be related through an experimentally obtained transfer coefficient, which describes the sensitivity of the resonator to changes in flux in the SQUID. 

We now examine the question of whether the resonator noise can be reduced below an assumed value of the SQUID noise. The approach would be to increase the flux-tuning range of the resonator, thus improving the flux sensitivity. Put simply, the more the SQUID tunes the resonator frequency, the less significant the resonator's resonant frequency jitter will be in comparison. If, for instance, the underlying SQUID noise is \SI{0.5}{\micro\fluxquanta\per\hertz\tothe{1/2}}, we estimate that a tuneability of \mbox{\SIrange{10}{20}{\mega\hertz}} would provide the necessary improvement in sensitivity. This could be achieved by including an array of rf-SQUIDs, or by reducing the inductance of the resonator by a factor of 2 while increasing the inductance of the non-junction arm of the SQUID by a factor of 5.

To conclude, we have presented the fabrication and characterisation of a Nb nanoSQUID-embedded coplanar resonator, containing a 3d nanobridge weak link. The Ne FIB fabrication of the nanobridge enables frequency tuneability at a temperature of $T = \SI{16}{\milli\kelvin}$, while maintaining a quality factor and intrinsic phase noise which is comparable to a bare CPW resonator fabricated on the same chip. The niobium material base is essential for compatibility with silicon spin systems, and the low temperature is critical for suppressing thermal occupation of excited states of the spin system. We believe that this flux-tuneable resonator can be an enabling technology for spin-based quantum memories. Using a homodyne detection scheme, we measured the flux sensitivity of the device, and outlined a recipe for future improvement. A modest increase in the flux tuning range should provide the necessary sensitivity to open up applications in high-sensitivity single-spin ESR.

\begin{acknowledgments}
Jamie A. Potter and Laith Meti contributed equally to this work.

Acknowledgement is given to the National Institute of Metrology, Beijing, China for providing the niobium films.

We acknowledge the support of the UK Department for Science, Innovation and Technology (DSIT) through the UK National Quantum Technologies Programme (NQTP), and the UK Science and Technology Facilities Council (STFC) through grants ST/T006064/1 and ST/T006099/1 - Quantum Sensors for the Hidden Sector (QSHS), and ST/T006137/1 - Quantum Technologies for Neutrino Mass (QTNM). We also acknowledge support through a UCL Impact Studentship.
\end{acknowledgments}

\bibliography{_references}

\end{document}